\documentclass[final]{aipproc}

\layoutstyle{6x9}

\newcommand{\graphicswidth}{0.5\textwidth}

\usepackage{amsmath,amssymb}

\begin{document}
\begin{flushright}
{\small
hep-ph/0610098\\
\hfill\\
October 9, 2006}
\end{flushright}
\vspace{-1cm}

\title{Top quark signatures of Higgsless models
\thanks{To appear in the proceedings of SUSY06, 
14th  International Conference on Supersymmetry 
and the Unification of Fundamental Interactions
Irvine, California, USA
12-17 June 2006 }
}

\author{Christian Schwinn}{
  address={ Institut f\"ur Theoretische Physik E, RWTH Aachen, 
    D - 52056 Aachen, Germany }
}

\begin{abstract}
In these proceedings I describe the use of tree level unitarity 
to constrain top quark signatures of Higgsless electroweak 
models and discuss implications for collider phenomenology.
\end{abstract}

\maketitle

\section{Higgsless electroweak symmetry breaking}

Recently 
higher dimensional Higgsless electroweak models 
have been proposed where 
the gauge symmetry is broken by boundary
conditions~\cite{Csaki:2003dt} and the bad high energy
behavior of $W$ and $Z$ scattering amplitudes is softened
by the exchange of $W$ and $Z$  Kaluza-Klein (KK)-modes instead of that of a
Higgs boson~\cite{Csaki:2003dt,SekharChivukula:2001hz,OS:SR}. 
Implementations of this idea employ e.g. 
a $SU(2)_L\otimes SU(2)_R\otimes U(1)$ 
in a warped extra
dimension~\cite{Csaki:2005vy,Burdman:2003ya}, a simple $SU(2)$
in a flat extra dimension~\cite{Foadi:2005hz} or four dimensional
$SU(2)^N\otimes U(1)^M$ 
`theory spaces'~\cite{Simmons:2006iw,Georgi:2005dm}.
Fermion masses in 5D Higgsless models
can be generated by gauge invariant brane localized mass and kinetic 
terms~\cite{Csaki:2005vy} without violating
the relevant unitarity sum rules (SRs)~\cite{CS:HLF}. 
Electroweak precision data can be satisfied by delocalizing
 the zero modes of light fermions 
and choosing appropriate bulk 
masses~\cite{Csaki:2005vy,Simmons:2006iw,Foadi:2005hz}.

Some signatures of Higgsless models---e.g. Drell-Yan production
of KK-gauge bosons---depend on details of the light fermion sector. 
In contrast, narrow KK-resonances below 1 TeV in vector boson
fusion~\cite{Birkedal:2004au} and anomalous gauge boson
 couplings~\cite{Csaki:2003dt,Simmons:2006iw} have
 been identified as a generic prediction of 
 the Higgsless mechanism.
We will briefly review the argument to prepare for an
analogous approach to top signatures~\cite{CS:HLTOP} 
in the next section.
Following~\cite{Birkedal:2004au}, one can use the unitarity sum rules  
ensuring the cancellation of terms
growing like $s^2$ and $s$ in  $WZ\to WZ$ scattering 
\begin{equation}
\label{eq:ww}  
g_{W^2Z^2}-g_{WWZ}^2 = \sum_n  g^2_{ZW W^{(n)}} \qquad
  g^2_{ZWW} m_Z^4= 3 m_W^2\sum_n  g^2_{ZWW^{(n)}} m^2_{W^{(n)}}
\end{equation}
 to estimate the coupling of the first $W$-KK-mode
as
$g_{ZWW^{(1)}}\lesssim (g m_Z)/(\sqrt 3 m_{W^{(1)}})$ which has found to be
promising for a detection at the LHC.
Since the Standard model~(SM) couplings alone satisfy the
first SR in~\eqref{eq:ww}, 
Higgsless models also
 require anomalous gauge boson couplings~\cite{Csaki:2003dt}.
Using~\eqref{eq:ww}, one estimates $ \delta g_{WWZ}/ g_{WWZ}=c\times
(m_W/m_{W^{(1)}})^2$ with $c\sim \mathcal{O}(1)$ in agreement with 
the result~\cite{Simmons:2006iw} $c= \pi^2/(12 \cos\theta_W^2)$ in a flat 5D
model.

\begin{figure}
   \includegraphics[width=0.475\textwidth]{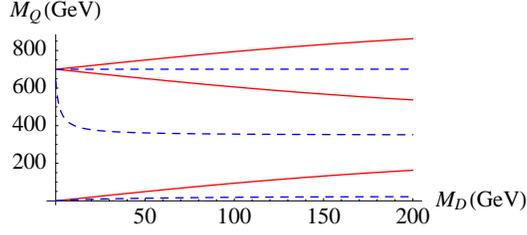}
 \caption{
Masses of bottom  (blue/dashed) and top (red/solid)
 zero- and KK-modes as function of the boundary mass parameter $M_D$ 
in a flat space  $SU(2)_L\otimes SU(2)_R\otimes U(1)$ toy model for 
$R^{-1}\!=700$ GeV}
\label{fig:kk}
\end{figure}

The large mass of the top quark implies a
special role for the third family.  
 As an example, figure~\ref{fig:kk} shows the mass spectrum in a flat space
$SU(2)_L\otimes SU(2)_R\otimes U(1)$ toy-model
 where the $t$ and $b$
quarks are embedded into bulk multiplets $Q_L=(t_L,b_R)$ and
$Q_R=(t_R,b_R)$ as in~\cite{Csaki:2005vy}. 
  One the brane at $y=\pi R$ the
symmetry breaking pattern $SU(2)_L\otimes SU(2)_R \to SU(2)_V$ allows
a localized mass term $\delta(y-\pi R) M_D(\bar Q_LQ_R+ \bar Q_RQ_L)$.
The large $M_D$ needed for the top mass splits the
 KK-modes of $Q_L$ and $Q_R$ leading to $m_{T^{(1)}}, m_{B^{(1)}}<m_{Z^{(1)}}$. Similar results are found in a warped model~\cite{Burdman:2003ya}.
For the embedding of $b_R$ chosen in this example, 
the bottom-top mass splitting requires a large boundary
kinetic term~(BKT) term for  $b_R$ at the $y=0$ brane where
$SU(2)_R$ is broken, further reducing $m_{B^{(1)}}$.
In such a setup, it has been found difficult to obtain a large enough
top quark mass while respecting bounds on anomalous $Z\bar b_Lb_L$ or
$W\bar t_Rb_R$ couplings~\cite{Burdman:2003ya,Csaki:2005vy,Foadi:2005hz}.  

Two strategies
have been proposed to overcome this difficulties:
\noindent a):  raising the mass of the $t$ and $b$ KK-modes, e.g. 
by introducing two slices of warped spaces with
different curvatures~\cite{Cacciapaglia:2005pa},  by
 allowing 5D Lorentz invariance
violating bulk-kinetic terms~\cite{Foadi:2005hz},  
or using theory space models~\cite{Georgi:2005dm}. 
\noindent b):
 protecting the $Zbb$ 
vertex by using a symmetric
representation 
of $SU(2)_L\otimes SU(2)_R$ for the $5D$ multiplet containing
$b_L$~\cite{Agashe:2006at,Cacciapaglia:2006gp}. 
Many third family signatures of Higgsless models 
depend on such model building choices, e.g. the strongest constraints
from anomalous couplings can arise from $Z \bar b_L b_L$~\cite{Csaki:2005vy} 
or $W\bar t_Rb_R$~\cite{Foadi:2005hz}
while large anomalous
 $Z\bar t t$ couplings arise in~\cite{Cacciapaglia:2006gp}.
The lower bound on $m_{T^{(1)}}$
ranges from $450$ GeV in~\cite{Cacciapaglia:2006gp}
to $1.6$ TeV~\cite{Foadi:2005hz}.
Finally, the two-bulk model of~\cite{Cacciapaglia:2005pa} features 
additional  scalar top-pions while the mechanism of~\cite{Agashe:2006at,Cacciapaglia:2006gp} involves KK-modes of colored fermions with $Q=\frac{5}{3}$.
\section{Unitarity constraints on top signatures}

As in the gauge boson sector~\cite{Birkedal:2004au},  
unitarity SRs can be used to constrain signatures of the Higgsless
mechanism in the top sector~\cite{CS:HLTOP}. 
In the 4D SM without a Higgs, the scattering amplitude
for $W^+W^-\to t\bar t$ grows like $\sqrt s$. 
In 5D Higgsless models 
the cancellation of this growth
by the exchange of $Z$ and bottom KK-modes
requires the SRs~\cite{CS:HLF,CS:HLTOP,Foadi:2005hz}
\begin{subequations}
\label{eq:wtb-srs}
 \begin{align}
\label{eq:tt-lie}
 -(g^{L/R}_{Wtb})^2+
     g_{ZWW}g_{ttZ}^{L/R}+ g_{WW\gamma}g_{tt\gamma}=
    \sum_n\left[(g^{L/R}_{WtB^{(n)}})^2
      -g_{WWZ^{(n)}}g_{ttZ^{(n)}}^{L/R}\right]=0\\
(g^{L}_{Wtb})^2+(g^{R}_{Wtb})^2= 
\sum_{n}\left[ 2\tfrac{m_{B^{(n)}}}{m_t}g^R_{WtB^{(n)}}g^L_{WtB^{(n)}}-(g^{L}_{WtB^{(n)}})^2-(g^{R}_{WtB^{(n)}})^2
 \right]
\label{eq:wtb-sr}
 \end{align}
\end{subequations}

Since the $Z$-KK-modes do not appear in~\eqref{eq:wtb-sr} , the
unitarity cancellations \emph{cannot be achieved} by including only
vector boson resonances and the presence of the third family quark
KK-modes is a generic consequence of the Higgsless mechanism.  The
left hand side of~\eqref{eq:tt-lie} vanishes in the SM, but this does
not necessitate anomalous top  couplings since in principle the
contributions of the $b$ and $Z$- KK modes can cancel each other.

To obtain estimates for the couplings, in~\cite{CS:HLTOP} the SRs from
 $W^+W^-\to t\bar t$, $WZ\to t b$ and $ZZ\to tt$ were solved under
 the assumption of saturation by the first KK-level and non-degenerate
 KK-modes.  For the example of equal left-and right handed
 couplings of the KK-modes and non-anomalous zero mode couplings
 one obtains the estimates
\begin{equation}\label{eq:wTb-hl}
\begin{aligned}
g_{Wt B^{(1)}}
&\approx \frac{g}{2}\sqrt{\frac{ m_t}{m_{B^{(1)}}}} ,&
   g_{tt Z^{(1)}}&
\approx \frac{\sqrt 3 g}{4}\frac{m_t m_{Z^{(1)}}}{m_W m_{B^{(1)}}} 
 ,& g_{t T^{(1)}Z}&\approx
 \frac{g}{2\cos\theta_w}\sqrt{\frac{m_t}{2m_{T^{(1)}}}} &&
\end{aligned}
\end{equation} 
We can check our assumptions in the 5D $SU(2)$
 model~\cite{Foadi:2005hz} where the top sector is parameterized by
 the coefficients $t_{L/R}$ of the BKTs
 and the 5D Lorentz violating mass term $\kappa\bar\Psi_t
 \partial_5 \Psi_t$. The
 flavor universal value $t_L\sim\pi m_W R/\sqrt 3$ ensures
 a vanishing $S$ parameter  while constraints on the $Wt_Rb_R$
vertex and unitarity of $W^+W^-\to t\bar t$ scattering
 imply the bound $3.6$ TeV$ <\kappa/R <32$ TeV~\cite{Foadi:2005hz}.  These
 parameters enter the top mass $m_t\approx\kappa t_Lt_{R}/(\pi R)$, 
the mass of the first bottom KK mode
 $m_{B^{(1)}}\approx \kappa/(2R)$ and its coupling constant $
 g^{L/R}_{WtB^{(1)}} \approx (2^{5/2} t_{L/R}/\pi^{2})  g^L_{Wtb}$.  
We find that the  
the SR~\eqref{eq:wtb-sr} is indeed saturated by the first KK-level
 to leading order in $t_{L/R}$.
 From  the bounds on $\kappa$  
we only obtain a weak constraint  
$ g^L_{WtB^{(1)}}/g^R_{WtB^{(1)}} \sim t_L/t_{R}
\sim (\kappa\pi R m_W^2)/(3m_t)\sim 0.2 - 2$ for $R^{-1}=800$ GeV.

\section{Implications for  phenomenology}
\begin{figure}[hbpt]
   \includegraphics[width=\graphicswidth]{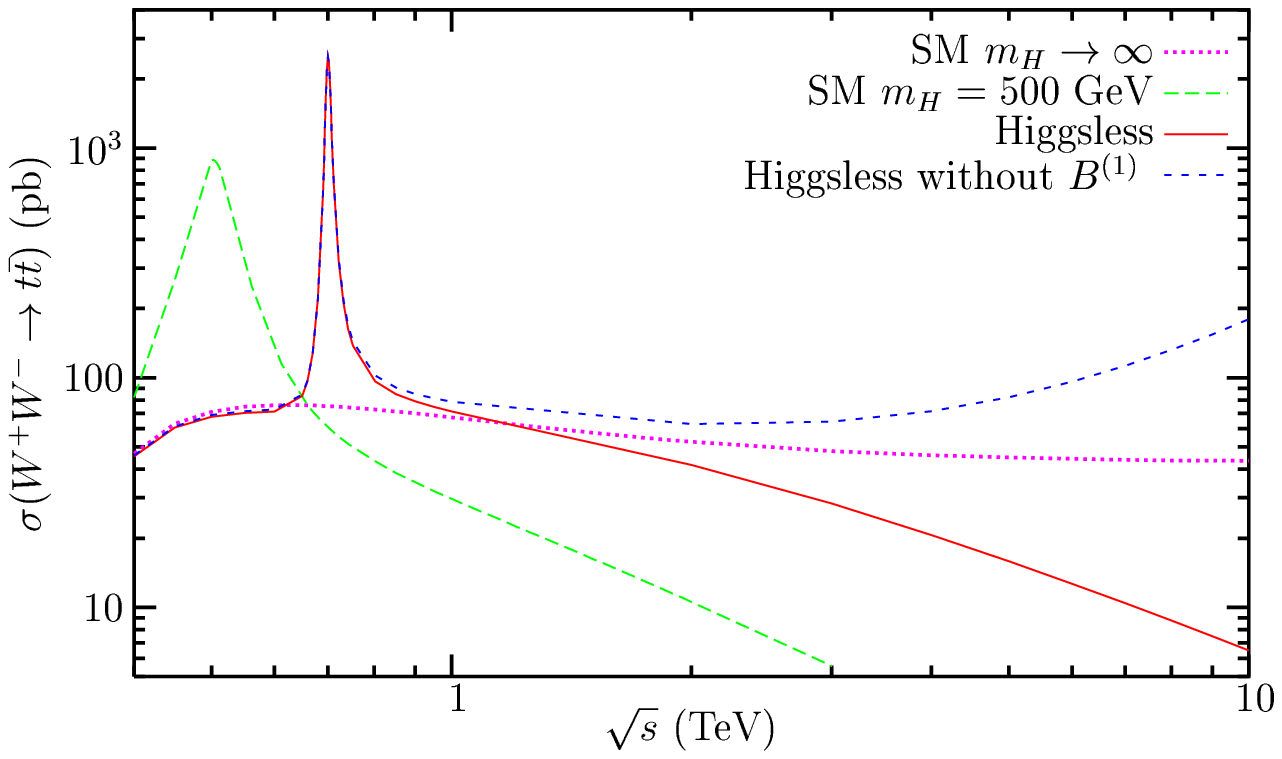}
   \includegraphics[width=\graphicswidth]{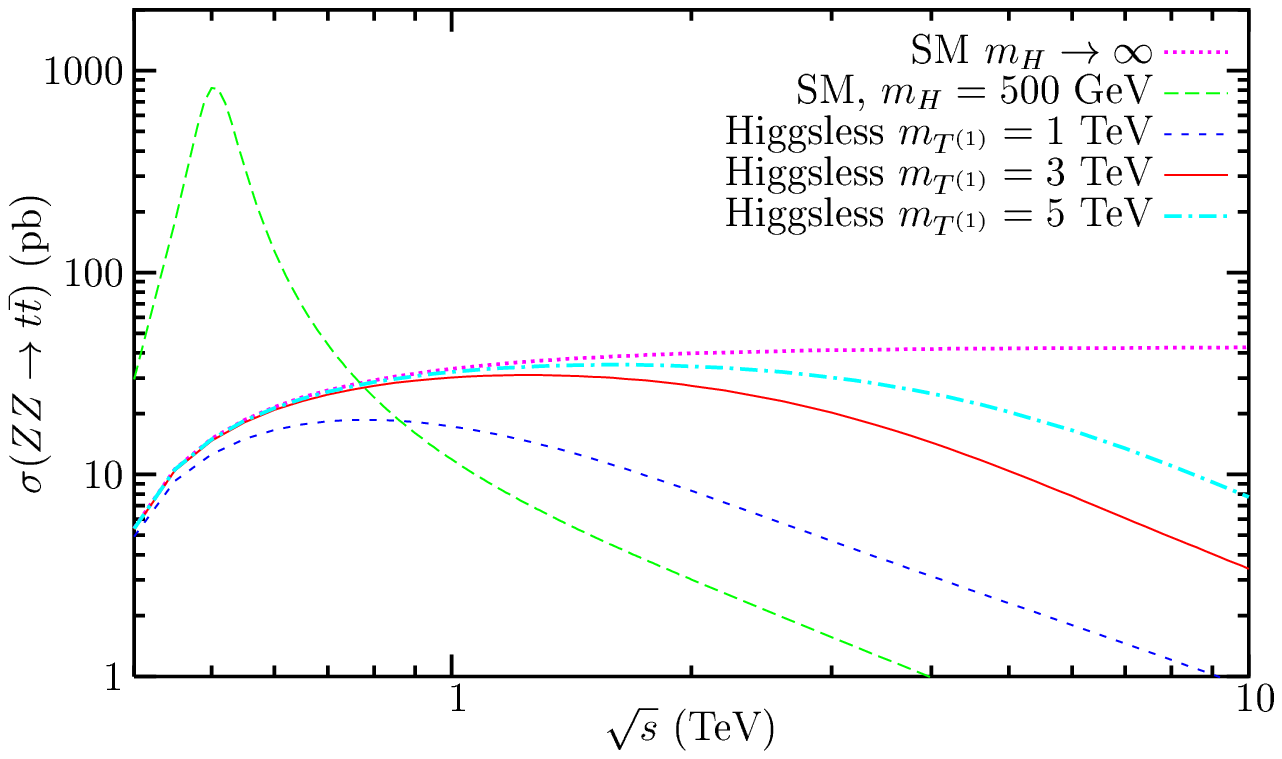}
    \caption{(left) Cross section for $W^+ W^-\to t\bar t$ (left)
      and  $ZZ\to t\bar t$  (right) in the
      SM with $m_H=500$ GeV, the SM with $m_H\to\infty$
      and the Higgsless scenario with $m_Z^{(1)}=700$ GeV and
      $m_{B^{(1)}}=2.5$ TeV 
      ~\cite{CS:HLTOP}. 
      }
\label{fig:wwtt}
\end{figure}

A simplified Higgsless scenario with a single KK-level and
 couplings as given in~\eqref{eq:wTb-hl} and  below~\eqref{eq:ww} can be used
to estimate collider signatures in the top sector.
The cross sections for $W^+ W^-\to t\bar t$ and $ZZ\to t\bar t$
displayed in figure~\ref{fig:wwtt} show a suppression at large $\sqrt
s$ both in the SM with a Higgs resonance and in the Higgsless scenario 
 while the inclusion of a
$Z^{(1)}$ without the $B^{(1)}$ rather \emph{destroys} the unitarity
cancellations present in the SM.  In  $ZZ\to t\bar t$ no
resonance appears in the Higgsless model so the unitarization is
entirely due to the $T^{(1)}$ in the $t$ channel and 
becomes effective earlier on for a lower $m_{T^{(1)}}$.

The couplings to third family quarks affect the phenomenology
of the gauge boson KK-modes. 
From~\eqref{eq:wTb-hl} the partial decay width of the $Z^{(1)}$
into top-quarks is estimated as
\begin{equation}
\label{eq:br_z1tt}
  \Gamma_{Z^{(1)}\to t\bar t}
\approx\frac{3\, \alpha_{QED} m^3_{Z^{(1)}}}{16 \sin^2\theta_w m_W^2}
 \left(\frac{m_t}{m_{B^{(1)}}}\right)^2
\approx 
\begin{cases}
11 \text{GeV} &(m_{B^{(1)}}=1 \text{TeV})\\
1.8\text{GeV}& (m_{B^{(1)}}=2.5 \text{TeV})
\end{cases}
\end{equation}
which is to be compared to 
$\Gamma_{Z^{(1)}\to WW}\approx 13$ GeV~\cite{Birkedal:2004au}
(here $m_{Z^{(1)}}=700$ GeV).

While a $Z^{(1)}$ resonance in $W^+ W^- \to \bar t t$ at the LHC is
overwhelmed by QCD background to $pp\to t\bar t jj$~\cite{Han:2003pu}, the
associated production with $b$ or $t$ quarks---e.g. in single top production
$W b\to t Z^{(1)}\to t t\bar t$ or top pair production $g g \to t\bar t
Z^{(1)}\to t\bar t t\bar t$---has been found promising for
 $m_{Z^{(1)}}=1$TeV~\cite{Han:2004zh},
however for a strong coupling to $b$ and $t$ with
 $\Gamma_{Z^{(1)}\to t \bar t+b\bar b}=
127$ GeV and negligible coupling to the $W$, unlike to Higgsless models.

The production of heavy top quarks in $W$-$b$ fusion $q b\to q' T$ 
at the LHC has
been studied in the context of Little Higgs models~\cite{Perelstein:2005ka}
 where
the relevant coupling 
 is given by $g_{WTb}^{L}=\frac{g}{\sqrt 2}
\frac{m_t\lambda_1}{m_T\lambda_2}$.
For $\lambda_1=\lambda_2$, the  reach of the LHC 
has been estimated~\cite{Perelstein:2005ka} as $m_T=2$ TeV. 
Comparing to the result analogous to~\eqref{eq:wTb-hl} 
for $g_{WT^{(1)}b}$~\cite{CS:HLTOP}
one can approximately relate  the cross sections in the Littlest
Higgs model and in the Higgsless scenario as
\begin{equation}
  \sigma_{\text{HL}}(Wb\to T^{(1)})=\frac{ m_b m_T^{(1)}}{m_t^2 }
 \left(\tfrac{\lambda_2}{\lambda_1}\right)^2  \sigma_{\text{LH}}(Wb\to T)
\approx 0.13 \frac{m_{T}^{(1)}}{\text{TeV}}
 \left(\tfrac{\lambda_2}{\lambda_1}\right)^2  \sigma_{\text{LH}}(Wb\to T)
\end{equation}
so  the phenomenology  of the $T^{(1)}$
will be even more challenging in the Higgsless model.
For a light $T^{(1)}$ as in~\cite{Cacciapaglia:2006gp}, QCD pair production is the dominant signal~\cite{Perelstein:2005ka}.

To summarize, 
viable Higgsless models exist~\cite{Foadi:2005hz,Georgi:2005dm,Cacciapaglia:2006gp}, but require a tuning of fermion 
mass parameters and a separate treatment of the third generation. 
Unitarity sum rules imply the presence of top and bottom KK-modes.
In addition,  associated production of the $Z^{(1)}$ with
top quarks deserves more detailed studies in  realistic models.


\begin{theacknowledgments}
The work described in this talk
was  supported by the DFG through the SFB/Tr 9 at RWTH Aachen
and the Graduiertenkolleg "Eichtheorien" at 
Mainz University.
\end{theacknowledgments}


\bibliographystyle{aipproc} 

\bibliography{biblio}

\end{document}